\documentclass[12pt]{article}
\usepackage{fullpage}
\usepackage{amsmath}
\usepackage{amssymb}
\usepackage{abstract}
\usepackage[hyperfootnotes=true,colorlinks=true,linkcolor=blue,citecolor=blue,urlcolor  = blue,anchorcolor = blue,linktocpage]{hyperref}
\numberwithin{equation}{section}
\usepackage[width=\textwidth]{caption}
\usepackage{cite} 

\newcommand{\be}{\begin{equation}}
\newcommand{\ee}{\end{equation}}

\newcommand{\bos}[1]{\boldsymbol{#1}}
\newcommand{\DD}[3]{D_{#1 #2}^{\;\;\;\;\; #3}} 
\newcommand{\cc}[3]{C_{#1 #2}^{\;\;\;\; #3}} 
\newcommand{\dd}[3]{D_{#1 #2}^{\;\;\;\; #3}}

\newtheorem{theorem}{Theorem}[section] 

\newtheorem{definition}{Definition}[section]

\title{Balancing Anisotropic Curvature with Gauge Fields\\ in a Class of Shear-Free Cosmological Models}

\date{\vspace{-5ex}}

\author{  
	Mikjel Thorsrud
	\\
	\small{\em Faculty of Engineering, \O stfold University College,}\\
	\small{\em P.O. Box 700, 1757 Halden, Norway.} \\
	\small{E-mail: mikjel.thorsrud@hiof.no }
}

\begin{document}

\maketitle

\begin{abstract}
	We present a complete list of general relativistic shear-free solutions in a class of anisotropic, spatially homogeneous and orthogonal cosmological models containing a collection of $n$ 
	independent $p$-form gauge fields, where $p\in\{0,1,2,3\}$, in addition to standard $\Lambda$CDM matter fields modelled as perfect fluids. 
	Here a (collection of) gauge field(s) balances anisotropic spatial curvature on the right-hand side of the shear propagation equation. The result is a class of solutions dynamically equivalent to standard FLRW cosmologies, with an effective curvature constant $K_{\rm{eff}}$ that depends both on spatial curvature and the energy density of the gauge field(s). In the case of a single gauge field ($n=1$) we show that the only spacetimes that admit such solutions are the LRS Bianchi type III, Bianchi type VI$_0$ and Kantowski-Sachs metric, which are dynamically equivalent to open ($K_{\rm{eff}}<0$), flat ($K_{\rm{eff}}=0$) and closed ($K_{\rm{eff}}>0$) FLRW models, respectively. With a collection of gauge fields ($n>1$) also Bianchi type II admits a shear-free solution ($K_{\rm{eff}}>0$). We identify the LRS Bianchi type III solution to be the unique shear-free solution with a gauge field Hamiltonian bounded from below in the entire class of models. 
\end{abstract}


\newpage
\tableofcontents
\newpage

\section{Introduction and main results \label{ch:intro}}

Available observational constraints on the dimensionless spatial curvature parameter $\Omega_K$ are restricted to the class of Friedmann-Lema\^{i}tre-Robertson-Walker (FLRW) metrics, i.e. spatially homogeneous and isotropic spacetimes. The most reliable constraint according to the Planck collaboration, combining cosmic microwave background (CMB) with baryon
acoustic oscillation, is $\Omega_K = 0.000 \pm 0.005 \; (95\%\; \rm{CL})$ \cite{int:Planck16}. As such, $\Omega_K$ is already dynamically negliglible as a source on the right-hand side of the Raychudhuri equation, affecting cosmic expansion by less than 1\% at any epoch post big bang nucleosynthesis. Although a window remains where it is possible and interesting to improve constraints, they are currently decaying towards a cosmic variance limit where $\Omega_K$ degenerates with large-scale metric perturbations \cite{int:Bull16,int:Waterhouse08}. 

It is natural to ask if such constraints would relax if the statistical analysis was carried out within the entire class of spatially homogeneous cosmological models, that include FLRW metrics as special cases. Although model dependent, the expected answer to this hypothetical question is \emph{typically no}. The reason is that spatial sections of Bianchi spacetimes generally are intrinsically anisotropic, with anisotropic curvature sourcing the shear propagation equation via the traceless three-dimensional Ricci tensor. The conformal expansion is therefore generally broken in such model, leading to anisotropic redshift of the CMB. This would produce large-scale $\Delta T / T$ fluctuations in the CMB not seen in observations \cite{int:Barrow85}. In fact, a 1\%-level spatial curvature in such models, would typically be accompied by a 1\%-level shear (see \cite{int:Thorsrud11} for examples of such tracking) resulting in a 1\%-level CMB temperature quadrupole \cite{int:Thorsrud12}, roughly two orders of magnitudes larger than the CMB quadrupole measured by Planck.

We emphasize that it is the shear, not the anisotropic curvature itself, that would produce large-scale anisotropies in the CMB incompatible with observations. In models where the conformal expansion is broken, the isotropy of the CMB (up to statistical fluctuations) observed today requires fine tuning in the initial conditions of the shear-tensor \cite{int:Lim99,int:Koivisto08, int:ThorsrudWithBarrow12}. The only possibilities for phenomenologically tractable cosmological models with anisotropic curvature, then seems to be either that the spatial curvature is extremely small (and consequently degenerated with large scale metric perturbations) or that the conformal expansion is intact. The latter is the case of interest in this paper, namely the case that the anisotropy is non-dynamical and purely intrinsic, i.e. restricted to three-dimensional hypersurfaces of homogeneity.

Theorems that employ the Einstein-Boltzmann equations to deduce implications betweeen cosmological models and symmetries of particle distributions, such as the Ehlers-Geren-Sachs theorem and its extensions \cite{int:Ehlers66,int:Ellis71,int:Stoeger94,bok:EllisMaartensMacCallum}, are sometimes used to claim that the observed (statistical) isotropy of matter distributions (such as the CMB) implies FLRW geometry. However, this clearly requires additional assumptions, including on symmetries of dark matter fields that we only observe indirectly via its coupling to gravitation. In fact, the existence of a conformal Killing vector field parallel to the four-velocity of comoving matter, which applies to the class of models investigated in this paper, guarantees for the isotropy of the background radiation \cite{int:Obukhov1992}; furthermore such spacetimes are known to be parallax-free \cite{int:Hasse88}.

The possibility of homogeneous, but anisotropic, shear-free cosmological models was first pointed out by Mimoso and Crawford \cite{main:Mimoso93}. They considered orthogonal models where the fluid flow is hypersurface orthogonal, geodesic and vorticity free, but in general not shear-free. They showed that shear-free evolution requires inclusion of an imperfect fluid that balances anisotropic curvature on the right-hand side of the shear propagation equation. Specifically, the shear-free condition reads $\pi_{\mu\nu}=2E_{\mu\nu}$, where $\pi_{\mu\nu}$ is the anisotropic stress of the  imperfect fluid and $E_{\mu\nu}$ is the electric Weyl tensor. This is a sufficient and necessary condition for the shear to remain zero in the considered class of models. This type of models were further investigated in \cite{main:Coley94,main:Coley94nr2}. The devolopment of cosmological perturbation theory for anisotropic cosmological models has been initiated in ~\cite{pert:Pereira07,pert:Uzan11,pert:nakamura11}.  Special attention has been devoted to perturbations in Kantowski-Sachs type shear-free models in \cite{pert:Pereira12} and \cite{pert:Pereira17}, the latter focusing on tensor modes.

Since the conformal expansion is intact in shear-free models, they provide an interesting phenomenology generalizing FLRW models. It is therefore interesting to find realisitic matter models capable of satisfying the shear-free condition. In the past shear-free solutions have been found only in a very special class of anisotropic spacetimes, namely Kantowski-Sachs type models where hypersurfaces of homogeneity are products between a flat direction and a maximally symmetric two-space. Carneiro and Marug\'an presented a solution in a locally rotationally symmetric (LRS) Bianchi type III spacetime realizing the shear-free condition with a massless scalar field \cite{main:Carneiro01}. Later Koivisto et. al. presented shear-free solutions balancing anisotropic curvature with a 2-form gauge field \cite{main:Koivisto11}, in the aforementioned LRS Bianchi type III metric as well as in the Kantowski-Sachs metric. As we shall see the shear-free LRS Bianchi type III solutions presented in \cite{main:Carneiro01} and \cite{main:Koivisto11} are in fact physically equivalent.

Note that the imperfect matter types employed to satisfy the shear-free condition in \cite{main:Carneiro01} and \cite{main:Koivisto11} both belong to the class of free $p$-form gauge theories \cite{bok:supergravity}. Specifically, the massless scalar field employed by Carneiro and Marug\'an \cite{main:Carneiro01} is the case $p=0$, whereas Koivisto et.al. \cite{main:Koivisto11} employed the case $p=2$. In this paper we systematically investigate the generality of this scenario within the class of homogeneous and orthogonal cosmological models. In such models all matter fields (including gauge fields) are comoving, i.e. have zero energy flux in the surfaces of homgoeneity. It follows that all matter fields are irrotational and non-accelerated.  We present a complete list of general relativistic shear-free solutions realized by a collection of $n$ independent free $p$-form gauge field, where $n\in \{ 1,2,3,\dots\}$ and $p\in\{0,1,2,3\}$.

\paragraph*{\textbf{Main results}} 
We seek to derive all general relativistic cosmological solutions for spacetimes with spatial sections $\Sigma_t$ that are
\begin{itemize}
	\item[A)] homogeneous and
	\item[B)] intrinsically anisotropic with a non-zero traceless three-dimensional Ricci tensor ${^3}\!S_{\mu\nu}$, 
	\item[C)] that expand isotropically, i.e. the congruence of fundamental observers is \emph{shear-free},	 
\end{itemize}
and that contain
\begin{itemize}
	\item[D)] a collection of matter fields modelled as comoving perfect fluids, accounting for standard $\Lambda$CDM constituents such as radiation, dust, dark matter and $\Lambda$, and
	\item[E)] a collection of $n$ independent $p$-form gauge fields with vanishing energy flux and action specified by (\ref{action}) in section \ref{ch:freepform}.
\end{itemize}

\begin{definition}[shear-free solution]
	Within our class of cosmological models a shear-free solution is a general relativistic solution that satisfies the list A) - E) of properties stated above.
	\label{def:shear-free}	
\end{definition}
Note that our definition, specifically point B), excludes trivial shear-free solutions with isotropic or vanishing spatial curvature, i.e. FLRW models. As a consequence, point E) is required in order to have a non-empty set of solutions.\footnote{Within the class of orthogonal and spatially homogeneous cosmological models containing only perfect fluids, the metric is FLRW iff the fluid flow is shear-free \cite{main:Collins82}.} All shear-free solutions covered by our definition belong to the class of orthogonal Bianchi models or the Kantowski-Sachs metric, the latter being the unique case of a spatially homogeneous spacetime that falls outside the Bianchi classification.
\; All shear-free solutions can be mapped onto standard FLRW cosmologies by employing an effective curvature constant $K_{\rm{eff}}$, defined in section \ref{ch:FLRWeq}, that depends both on spatial curvature ($^3\!R$) and the total energy density of the gauge fields ($\rho_\mathcal{A}$). The shear-free solutions are dynamically equivalent to open, flat and closed FLRW models when $K_{\rm{eff}}$ is negative, zero and positive, respectively. 

A last technicality that appears in the summary of main results below is the symmetric tensor density $N^{ab}$, whose eigenvalues is used to classify all Bianchi models into 9 invariant types enumerated from I to IX (see section \ref{ch:BianchiTypes}). Among some of the Bianchi types there is an invariant subtype that satisfies the frame invariant condition $N^a_{\;\;a}=0$, which we identify to be an important subtype among shear-free solutions. 

In order to guarantee that the solutions we derive are the most general, our metric ansatz is the most general one compatible with the considered class of models (with 6 parameters of the spatial metric describing off-diagonal as well as diagonal components). Our main results can be summarized in table \ref{tab:shearfree} and the following three theorems.
\begin{table}[h]
	\newcommand\T{\rule{0pt}{2.6ex}}
	\newcommand\B{\rule[-1.2ex]{0pt}{0pt}}
	\begin{center}
		\begin{tabular}{ l c c c c c c c c}
			\hline\hline 
			Spacetime (subtype) \T     &  \footnotesize{sign $K_{\rm{eff}}$} & \footnotesize{ sign $^3\!R$} &  $n_{\rm{min}}$ & \footnotesize{Hamiltonian bounded} &    \\ 
			\B    &   &  &       & \footnotesize{from below?}     \\\hline
			\T\B Bianchi type II & $+$  & $-$  & 2 & no  \\
			\T\B Bianchi type VI$_0$ ($N^a_{\;\;a}=0$) &  $0$  & $-$  & 1 & no  \\
			\T\B Bianchi type III ($N^a_{\;\;a}=0$, LRS) &  $-$  & $-$    & 1 & yes  \\
			\T\B Kantowski-Sachs &  $+$  & $+$    & 1 & no  \\ \hline
			\T\B FLRW (closed/flat/open) &  $+/0/-$  & $+/0/-$ &       & \\ 
			\hline\hline
		\end{tabular}
		\caption{Complete list of shear-free solutions and comparison with FLRW standard cosmologies. Here $K_{\rm{eff}}$ is the effective curvature constant defined in (\ref{keff}), $^3\!R$ is the three-dimensional Ricci scalar on hypersurfaces of homogeneity and $n_{\rm{min}}$ is the minimum number of independent gauge fields required by the shear-free condition.} 
		\label{tab:shearfree}
	\end{center}
\end{table}

\begin{theorem}[a single gauge field]
	The only spacetimes that admit shear-free solutions (definition \ref{def:shear-free}) realized with a single gauge field ($n=1$) are the LRS Bianchi type III ($N^a_{\;\;a}=0$), Bianchi type VI$_0$ ($N^a_{\;\;a}=0$) and the Kantowski-Sachs metric, which have expansion histories equivalent to open, flat and closed FLRW models, respectively.  
	\label{int:theorem1}	
\end{theorem}
In each of the shear-free solutions covered by Theorem \ref{int:theorem1} the shear-free condition is realized by a $p$-form gauge field with $p\in\{0,2\}$ because, as we show, the cases $p\in\{1,3\}$ are not capable of balancing anisotropic curvature. The cases $p=0$ and $p=2$ are physically equivalent via Hodge dual at the field strength $p+1$ level, as shown in section \ref{ch:freepform}. We employ a unified description for the two cases $p=0$ and $p=2$ and thus there are two formal copies of each shear-free solution covered by Theorem \ref{int:theorem1}. This unifies the LRS Bianchi type III shear free solutions realized by $p=0$ in \cite{main:Carneiro01} and with $p=2$ in \cite{main:Koivisto11}, which are physically equivalent. 

Due to constraint equations, the field strength of each gauge field has only one independent degree of freedom in the shear-free solutions covered by Theorem \ref{int:theorem1}. The generalization to multiple gauge fields is therefore rather trivial for these spacetimes, since the field strengths are aligned and span a line. However, when we consider multiple gauge fields ($n>1$) a new type of shear-free solution appears in Bianchi type II.
\begin{theorem}[multiple gauge fields]
	The only spacetimes that admit shear-free solutions (definition \ref{def:shear-free}) realized by a collection of gauge fields ($n>1$) are the Bianchi type II, LRS Bianchi type III ($N^a_{\;\;a}=0$), Bianchi type VI$_0$ ($N^a_{\;\;a}=0$) and the Kantowski-Sachs metric, which have expansion histories equivalent to closed, open, flat and closed FLRW models, respectively.  
	\label{int:theorem2}	
\end{theorem}
In order to establish this theorem a careful analysis of Bianchi type II is required, which is the unique Bianchi model with a non-Abelian Lie algebra where the field strengths span a plane. A cruical step in our analysis is the identification of an automorphism, a gauge transformation of frame that preserves the Lie-algebra, which is used to identify the true degrees of freedom upon gauge fixing. 

In this paper we derive all general relativistic solutions without assuming any energy condition from the start. Based on an equivalence between the weak energy condition and the boundedness of the Hamiltonian, established in section \ref{ch:Hamiltonian}, we introduce a simple procedure for classifying solutions according to the boundedness of the Hamiltonian in section \ref{ch:shearfreecondition}. This results in the following theorem:
\begin{theorem}[boundedness of the Hamiltonian]
	The LRS Bianchi type III ($N^a_{\;\;a}=0$) is the unique spacetime that admits shear-free solutions (definition \ref{def:shear-free}) where the Hamiltonian of each gauge field is bounded from below. 
	\label{int:theorem3}	
\end{theorem}

Despite the fundamental importance of boundedness of the Hamiltonian, the full list of shear-free solutions is phenomenologically interesting; it provides examples on new types of behavior that can occur in this type of models. Note that the correspondence $K_{\rm{eff}} \gtrless 0 \iff {^3}\!R \gtrless 0 $ is broken by two new shear-free solutions of Bianchi type II and VI$_0$. Curiously, the latter has negative curvature (${^3}\! R <0$), but is dynamically equivalent to a flat FLRW model ($K_{\rm{eff}}=0$).


\paragraph*{\textbf{Outlook}}

It is possible to extend our knowledge of shear-free cosmological models in numerous ways, for instance, by considering other matter candidates, other cosmological models or other theories of gravitation.\footnote{See \cite{shear-free:Abebe16} for an investigation of shear-free cosmological models in $f(R)$ gravity. See \cite{main:Coley94} and \cite{main:Coley94nr2} for investigations of other cosmological models.} It is expected that several of the strategies developed in this paper can be utilized more or less directly in such investigations. This includes the framework introduced in section \ref{ch:shearfreecondition}, as well as our treatment of Bianchi type II, which demonstrates how automorphisms can be utilized to simplify the analysis of models with imperfect fluids, with no loss of generality. 


\paragraph*{\textbf{Organization of paper}}
In section \ref{ch:freepform} we investigate the familiy of gauge field theories with a free $p$-form, paying special attention to the structure of the energy-momentum tensors. 
In section \ref{ch:balancing} we review shear-free orthogonal models, show that a Maxwell vector ($p=1$) is not capable of balancing anisotropic curvature in such models and demonstrate the dynamical equivalence of our considered class of models with standard FLRW models.  In section \ref{ch:main:class&con} the Bianchi classification is introduced and employed to identify the degrees of freedom of the field strength in all Bianchi models. In section \ref{ch:shearfree} we first develop a framework for deriving general relativistic shear-free solutions. Based on this framework, we systematically derive and classify all shear-free solutions in the following subsections.

\paragraph*{\textbf{Conventions and notation}} Throughout we use Misner, Thorn and Wheeler's sign conventions for the metric tensor, Riemann tensor and Ricci tensor \cite{bok:MTW}. Greek letters ($\alpha$, $\beta, \dots$) are used for spacetime indices and run from 0 to 3. Latin indices ($a$, $b, \dots$) run from 1 to 3 and are employed to describe three-dimensional spatial hypersurfaces as well as Lie algebras. We use units in which $8\pi G=1$ and $c=1$. Throughout, $\boldsymbol{\mathcal A}$ denotes a generic $p$-form gauge field and $\boldsymbol{\mathcal  F}$ denote the associated $p+1$ field strength.

\section{Free p-form gauge theories \label{ch:freepform}}

To design shear-free Bianchi cosmologies a matter source capable of balancing anisotropic spatial curvature on the right-hand side of the shear propagation equation is needed. A natural candidate is the class of free theories described by the action
\be 
S = -\frac{1}{2} \int \boldsymbol{ \mathcal F} \wedge  \star \boldsymbol{ \mathcal F} \,, 
\ee
where $\bos{\mathcal  F} = \boldsymbol{\mathrm d \mathcal A}$ is the field strength of a $p$-form gauge field 
\begin{equation}
\boldsymbol{\mathcal A} = \frac{1}{p!} \mathcal A_{\mu_1\dots \mu_p} \mathbf w^{\mu_1} \wedge  \dots \wedge  \mathbf w^{\mu_p}\,.
\end{equation}
Here $\{ \mathbf w^\mu \}$ is an arbitrary one-form basis dual to $\{\mathbf e_\mu \}$. In terms of components the action can be written
\be
S = -\frac{1}{2(p+1)!} \int d^4x \sqrt{-g} \mathcal F^{\mu_1 \dots \mu_{p+1}} \mathcal F_{\mu_1 \dots \mu_{p+1}}\,,
\label{action}
\ee
where $\mathcal F_{\mu_1 \dots \mu_{p+1}} = (p+1) \nabla_{[\mu_1} \mathcal A_{\mu_2 \dots \mu_{p+1}]}$. The components of the energy-momentum tensor are
\be
T_{\mu\nu} \equiv - \frac{2}{\sqrt{-g}} \frac{\delta \mathcal L}{\delta g^{\mu\nu}} = \frac{1}{p!} \mathcal F_\mu^{\;\; \alpha_1 \dots \alpha_p} \mathcal F_{\nu\alpha_1\dots \alpha_{p}} - \frac{1}{2(p+1)!} g_{\mu\nu}
\mathcal F^{\alpha_1 \dots \alpha_{p+1}} \mathcal F_{\alpha_1\dots \alpha_{p+1}} \,.
\label{emt:general}
\ee

We shall consider all non-trivial $p$-forms in spacetime described by this action, i.e. $p\in\{0,1,2,3\}$. However, since our goal is to present a complete list of exact shear-free solutions according to definition \ref{def:shear-free}, we can already rule out the case $p=3$ which is equivalent to the cosmological constant $\Lambda$ (and consequently exluded by point B of definition \ref{def:shear-free}) \cite{int:Hawking84}. 
Thus, our attention will be restricted to $p\in\{0,1,2\}$ from here on. 

For each gauge field the closedness of the field strength gives a Bianchi identity
\be \boldsymbol{\mathrm d \mathcal F} = 0\,, \ee
whereas the equation of motion obtained by calculus of variations can be written 
\be \boldsymbol{\mathrm d \star \mathcal F} = 0\,. \ee
Notice the symmetry in these equations under the Hodge dual $\bos{\mathcal F}\rightarrow \star\bos{\mathcal F}$.
Below we shall see that this symmetry also holds for the energy-momentum tensor. In fact the theory of a canonical $p$-form with action (\ref{action}) is \emph{physically equivalent} to the similar  ($2-p$)-form theory via the Hodge dual at the field strength $p+1$ level \cite{bok:supergravity}. The details of this duality between a 2-form gauge field and a 0-form gauge field (a canonical massless scalar) is considered below.  For the Maxwell case $p=1$ the symmetry is the well-known self-duality in which electric and magnetic components transform into each others. 

\paragraph*{\textbf{1+3 covariant decompostion}} 

For each $p$ we shall decompose the energy-momentum tensor (\ref{emt:general}) according to the standard $1+3$ covariant decomposition
\be
T_{\mu\nu}=\rho u_\mu u_\nu+ P  h_{\mu\nu}+2q_{(\mu}u_{\nu)}+\pi_{\mu\nu}\,,
\ee
where $h_{\mu\nu}=g_{\mu\nu}+u_\mu u_\nu$ is the local metric of the instantaneous rest space orthogonal to the timelike unit norm vector field $u^\mu$. From section \ref{ch:balancing} we will identify $u^\mu$ with the four-velcocity of a \emph{comoving observer}, in which case $h_{\mu\nu}$ is the induced metric on hypersurfaces of homogeneity. Until then $u^\mu$ is identified with the four-velocity of an arbitrary observer, that sees the following energy density $\rho$, pressure $P$, energy flux $q^\gamma$ and anisotropic stress $\pi_{\mu\nu}$:
\be
\rho=u^\mu u^\nu T_{\mu\nu}\,,\;
P = \frac{1}{3}h^{\mu\nu}T_{\mu\nu}\,,\;
q^{\gamma} = -h^{\gamma \mu} u^\nu T_{\mu\nu}\,, \;\
\pi_{\mu\nu} = \left( h_{(\mu}^{\;\;\;\alpha} h_{\nu)}^{\;\;\;\beta} - \frac{1}{3}h_{\mu\nu}h^{\alpha\beta} \right) T_{\alpha\beta}\,.
\ee
We notice that the last two observables belong to the space orthogonal to the observer: $0=u^\gamma q_\gamma=u^\mu\pi_{\mu\gamma}$. Since the anisotropic stress is traceless, $\pi^{\mu}_{\;\;\mu}=0$, we sometimes refer to it as anisotropic pressure, whereas the tracepart goes into the manifestly isotropic pressure $P$.

\subsection{0- and 2-form gauge fields and their equivalence \label{ch:0-2-form}}

As mentioned above, the action (\ref{action}) constructed from gauge fields with $p=0$ and $p=2$ are physically equivalent. For these two cases we use the notation $\boldsymbol{\mathcal A} = \{ \phi, \boldsymbol{B} \}$ and $\boldsymbol{\mathcal F} \equiv \boldsymbol{\mathrm d \mathcal A} = \{ \mathbf \Phi, \boldsymbol{J} \}$ for $p=0$ and $p=2$, respectively. In order to obtain a unified notation for the physically equivalent fields we introduce a 1-form $\boldsymbol X$ which represents $\mathbf \Phi$ if $p=0$ and $\star \boldsymbol J$ if $p=2$. In component notation:  
\be
X_\mu = 
\begin{cases} 
	\nabla_\mu \phi,  & \text{ if }\; p=0 \,, \\
	\ast J_\mu, &  \text{ if }\; p=2 \,,            
\end{cases}
\label{def:eq:X}
\ee
where $\ast \!J_\delta = \frac{1}{6} \eta_{\alpha\beta\gamma\delta} J^{\alpha\beta\gamma}$. We shall refer to $X_\mu$ as the \emph{field strength}, although this is literally correct only for the case $p=0$.
The equations of motion, Bianchi identity and energy-momentum tensor can then be written as
\begin{align}
\nabla_\mu X_\nu - \nabla_\nu X_\mu = 0& \quad  \leftrightarrow \quad  \bos{\mathrm d X} = 0 \,, \label{eq:X1}  \\
\nabla_\gamma X^\gamma = 0& \quad  \leftrightarrow \quad \bos{\mathrm d \star X} = 0 \,, \label{eq:X2}  \\
T_{\mu\nu} = X_\mu X_\nu -\frac{1}{2} g_{\mu\nu} X_\gamma  X^\gamma\,.   &  
\label{F3}
\end{align}
Note the duality under $\Phi_\mu \leftrightarrow \ast J_\mu$ in which the energy-momentum tensors transform into each others and the Bianchi identity of either ($\boldsymbol{\mathrm d \mathcal F}=0$) transforms into the equation of motion of the other ($\boldsymbol{\mathrm d \star \mathcal F}=0$). Since the field strength is the only observable in the free theories defined by the action (\ref{action}), this is enough to establish the physical equivalence between the the cases $p=0$ and $p=2$. 

\paragraph*{\textbf{1+3 covariant decompostion}} We decompose the Hodge dual of the 3-form field strength as 
\be
X_\mu = -\varphi u_\mu + v_\mu,
\ee
where $v^\alpha$ is a spacelike vector orthogonal to $u^\alpha$, i.e. $v^2=v^\gamma v_\gamma>0$ and $u^\gamma v_\gamma=0$. The energy-momentum tensor (\ref{F3}) can then be decomposed as
\be
\rho =\frac{1}{2} (v^2 + \varphi^2)\,,  \quad
P = \frac{1}{2} \left(-\frac{1}{3}v^2+\varphi^2\right)\,,\quad
q^\mu =-\varphi v^\mu\,, \quad 
\pi_{\mu\nu} = v_\mu v_\nu - \frac{1}{3} v^2 h_{\mu\nu} \,. 
\label{siste1}
\ee
Notice that the energy flux $q^\mu$ vanishes only if $\boldsymbol X$ is purely spacelike ($X_\mu = v_\mu$) or purely timelike ($X_\mu = -\varphi u_\mu$) relative to the observer. Contrary to the Maxwell case $p=1$, considered below, the equation of state $P/\rho$ is a dynamical entity that depends both on the field ($\boldsymbol X$) and the observer ($\mathbf u$). We note that the 2-form has equation of state $P/\rho=-1/3$ if $\boldsymbol X$ is orthogonal to the observer. In section \ref{ch:balancing} we shall see that this particular equation of state is cruical in order to balance anisotropic curvature on the right-hand side of the shear propagation equation.

\subsection{1-form gauge field \label{ch:1-form}}
The usual Maxwell Lagrangian 
\be
\mathcal L = -\frac{1}{4} \sqrt{-g} F^{\mu\nu}F_{\mu\nu} 
\ee
gives:
\begin{align}
\nabla_\alpha F_{\beta\gamma} + \nabla_\beta F_{\gamma\alpha} + \nabla_\gamma  F_{\alpha\beta} = 0&  \quad \leftrightarrow \quad  \bos{\mathrm d F} = 0 \,,  \\
\nabla_\alpha F^{\alpha \gamma}=0& \quad \leftrightarrow \quad \bos{\mathrm d \star \!F} = 0 \,, \\
T_{\mu\nu} = F_{\mu \gamma}F^{\;\;\gamma}_{\nu} - \frac{1}{4} g_{\mu\nu} F^{\alpha\beta}F_{\alpha\beta} \,.
\label{Fv0}
\end{align}

\paragraph*{\textbf{1+3 covariant decompostion}} Upon introducing electric or and magnetic components relative to the observer $u^\mu$
\be
 E_\mu = F_{\mu\nu} u^\nu, \quad B_\mu = \ast F_{\mu\nu} u^\nu = \frac{1}{2} \eta_{\mu\alpha\beta} F^{\alpha\beta} \iff \quad F_{\mu\nu} = 2u_{[\mu}E_{\nu]} + \eta_{\mu\nu\alpha} B^\alpha \,,
\ee
the energy-momentum tensor (\ref{Fv0}) can be decomposed as
\begin{align}
&\rho = \frac{1}{2} (E^2+B^2) \,,  \quad
P = \frac{1}{6} (E^2+B^2)\,,  \quad
q^\mu = \eta^{\mu\alpha\beta}  E_\alpha B_\beta\,, \label{ny:1}  \\
&\pi_{\mu\nu} = -E_\mu E_\nu + \frac{1}{3} E^2 h_{\mu\nu} -B_\mu B_\nu + \frac{1}{3} B^2 h_{\mu\nu} \,. \label{ny:2}
\end{align}
Notice that the scalars $\rho$ and $P$ are invariants only in the covariant sense, i.e. the energy density and pressure seen by the observer depends not on our choice of frame, but of course do depend on the four-velocity $u^\mu$ of the observer. However, $P/ \rho=1/3$ for all electromagnetic fields and for all observers. Therefore, the equation of state is an invariant in a stronger sense; it is independent of $F_{\mu\nu}$ and a Lorentz invariant (independent of the observer $u^\mu$). We noted in the section above that $T_{\mu\nu}$ of gauge fields with $p=\{0,2\}$ has quite different properties, with a dynamical equation of state that depends on the four-velocity of the observer.

\subsection{Boundedness of the Hamiltonian \label{ch:Hamiltonian}}

General relativity does not enforce energy conditions by itself, and additional conditions are typically based on quantum theory, such as boundedness from below of the Hamiltonian which is necessary for a stable vacuum state \cite{int:vector-theories}. We will not assume any energy condition in the matter sector from the start in this work, but in the end we are of course interested in identifying physically viable solutions.  In this section an equivalence between the weak energy condition and the boundedness of the Hamiltonian is established, for our family of gauge theories. In section \ref{ch:shearfreecondition} this result will be used to establish a procedure for classifying shear-free solutions according to the boundedness of the Hamiltonian.

The weak energy condition stipulates that \cite{bok:Hawking} 
\be
T_{\mu\nu} V^\mu V^\nu \ge 0 
\ee
for all timelike vector fields $\mathbf V$. Physically this means that all observers see a positive energy density.  Since our only assumption about the timelike vector $u^\mu$ in the 1+3 covariant decomposition of $T_{\mu\nu}$ above was unit norm, the expressions for $\rho$ calculated in (\ref{siste1}) and (\ref{ny:1}) are sufficient for proving that the weak-energy condition holds for the class of gauge theories defined by the action (\ref{action}).

Next we turn our attention to the Hamiltonian density 
\be
\mathcal H \equiv \frac{\partial \mathcal L}{\partial(\partial_0 \mathcal A_{\mu_1\dots \mu_p})} \partial_0 \mathcal A_{\mu_1\dots \mu_p} - \mathcal L.
\ee
In a local Lorentz frame where $\nabla_\mu = \partial_\mu$ and $g_{\mu\nu}= \rm{diag}(-1,1,1,1)$ it takes the form 
\be
 \mathcal H = \frac{1}{p!} \mathcal F_{0}^{\;\; a_1 \dots a_p} \mathcal F_{0 a_1\dots a_{p}} + \frac{1}{2(p+1)!}
\mathcal F^{\alpha_1 \dots \alpha_{p+1}} \mathcal F_{\alpha_1\dots \alpha_{p+1}} + \frac{p}{p!} \mathcal F^{b_1\dots b_p 0} \partial_{b_1} \mathcal A_{b_2\dots b_p 0}\,,
\label{hamiltonian}
\ee 
which represents the energy component of the canonical energy-momentum tensor, i.e. the conserved current under infinitesimal coordinate transformations. This tensor is not identical to the energy-momentum tensor (\ref{emt:general}) obtained by variation with respect to the metric, as seen clearly by the lack of gauge invariance in (\ref{hamiltonian}). 

Notice that the Hamiltonian density (\ref{hamiltonian}) agrees with the energy density
\be
\rho = u^\mu u^\nu T_{\mu\nu}
\ee
seen by the observer $u^\mu=(1,0,0,0)$  only up to the third term in $\mathcal H$. This term represents a well-known ``defect'' of the canonical energy-momentum tensor ($\mathcal T^{\mu\nu}$), which is neither symmetric nor gauge invariant in general \cite{ny:emt:canonical}. This can be cured in the standard way \cite{ny:Landau} by adding the divergence of a Belifante tensor $\mathcal T^{\mu\nu} \rightarrow \mathcal T^{\mu\nu} + \partial_\lambda K^{\mu\lambda\nu}$, which satisfies $K^{\mu\lambda\nu}=-K^{\lambda\mu\nu}$. By choosing 
\be
K^{\mu\lambda\nu}=\frac{p}{p!} \mathcal F^{\mu\lambda\alpha_1\dots \alpha_{p-1}} \mathcal A^\nu_{\;\;\alpha_1\dots \alpha_{p-1}}
\ee
the gauge dependent term in (\ref{hamiltonian}) vanishes and $\mathcal T^{\mu\nu}$ becomes symmetric. The equivalence between the weak energy condition and the boundedness of the Hamiltonian follows. 

\section{Balancing anisotropic curvature with a comoving gauge field \label{ch:balancing}}

\subsection{Shear-free condition \label{ch:orthogonal}}
Bianchi spacetimes admit a slicing in three-dimensional spacelike hypersurfaces $\Sigma_t$, each labeled by the time parameter $t=x^0$, with constant curvature scalars. We let $u^\mu$ denote the unique timelike vector field ($u^\mu u_\mu=-1$) that is orthogonal to such surfaces of homogeneity.  We are interested in \emph{orthogonal} Bianchi models in which $u^\mu$ also represent the flow of all matter fields. In such models the congruence of fundamental observers are non-accelerated and irrotational and the associated kinematically relevant quantities are the Hubble expansion scalar
\be
3H = \nabla_\mu u^\mu
\ee
and the shear tensor
\be
\sigma_{\mu\nu} = \nabla_{(\nu} u_{\mu)} - H h_{\mu\nu} \,.
\ee
Our cosmological model contains a collection of comoving perfect matter fields, for instance $\Lambda$CDM constituents, with total energy density $\rho$ and total pressure $P$. We also consider a $p$-form gauge field $\mathcal A_{\mu_1\dots\mu_p}$ with zero energy flux in $\Sigma_t$ (thus referred to as ``comoving''), with energy-density $\rho_{\mathcal A}$, pressure $P_{\mathcal A}$ and anisotropic stress $\pi_{\mu\nu}$ (which are total quantities in the case of multiple gauge fields). The total energy-momentum tensor is thus  
\be
T_{\mu\nu} = (\rho_{\mathcal A} + \rho ) \; u_\mu u_\nu + (P_{\mathcal A}+P) \; h_{\mu\nu} + \pi_{\mu\nu} \,.  
\label{totalEMT}
\ee
The matter fields satisfy the evolution equations
\begin{align}
&\dot \rho + 3H(\rho+P) = 0 \,, \label{eq:evPF} \\
&\dot \rho_{\mathcal A} + 3H(\rho_{\mathcal A}+P_{\mathcal A}) = -\pi^{\mu\nu} \sigma_{\mu\nu} \label{eq:evA} \,,
\end{align}
whereas the congruence evolves according to the Raychudhuri equation and the shear propagation equation
\begin{align}
&\dot H + H^2 = -\frac{1}{6} (\rho + 3P) -\frac{1}{6} (\rho_\mathcal{A} + 3P_\mathcal{A}) - \frac{2}{3} \sigma^2 \,, \label{eq:raychudhuri} \\ 
&\dot \sigma_{\mu\nu} + 3H\sigma_{\mu\nu} = \pi_{\mu\nu} - {^3}\!S_{\mu\nu} \,.
\label{eq:shear}
\end{align}
Here an overdot denote a time derivative along the congruence, i.e. $\dot \rho = u^\mu \nabla_\mu \rho$, and
\be ^3\!S_{\mu\nu} = {^3}\!R_{\mu\nu} - \frac{{^3}\!R}{3} h_{\mu\nu} \ee
is the trace-free three-dimensional Ricci tensor on the hypersurfaces $\Sigma_t$. There is one Hamiltonian constraint among the variables 
\be
3H^2 - \sigma^2 + \frac{^3\!R}{2}  = \rho + \rho_{\mathcal A}\,,
\label{eq:friedmann}
\ee
that we will sometimes refer to as the ``Friedmann equation''. To write the shear propagation equation (\ref{eq:shear}) on a simple form we replaced the electric components of the Weyl tensor
\be
E_{\mu\nu} = C_{\mu\alpha\nu\beta} u^\alpha u^\beta
\ee
(where $C_{\mu\alpha\nu\beta}$ is the Weyl tensor) using the relation
\be
{^3}\!S_{\mu\nu} = E_{\mu\nu} + \frac{1}{2} \pi_{\mu\nu} - H \sigma_{\mu\nu} + \sigma_{\mu}^{\;\;\gamma} \sigma_{\gamma\nu} - \frac{2}{3} \sigma^2 h_{\mu\nu}\,,
\label{relation}
\ee
which is valid for irrotational congruences \cite{bok:EllisWainwright}. 

Following Mimoso and Crawford \cite{main:Mimoso93}, we consider (\ref{eq:shear}) with $\sigma_{\mu\nu}=0$ and realize that
\be
\pi_{\mu\nu} = {^3}\!S_{\mu\nu}, 
\label{shearfree:S}
\ee
or equivalently
\be
\pi_{\mu\nu} = 2E_{\mu\nu}\,,
\label{shear-free-con}
\ee
is a necessary and sufficient condition for the shear to remain zero. In this paper we shall investigate systematically under which conditions gauge fields are capable of balancing the anisotropic curvature in this way. 
\subsection{Excluding a free Maxwell type gauge field \label{ch:candidates}}

We are now ready to show which type of $p$-forms described by the action (\ref{action}) are capable of balancing anisotropic curvature by satisfying the shear-free condition (\ref{shearfree:S}). Based solely on the structure of the energy-momentum tensor we will rule out the Maxwell case $p=1$, which is the main result of this section. Naively, the reason is simply that in the shear-free limit the energy density of a Maxwell field decays adiabatically as $1/a^4$ whereas anisotropic curvature decays as $1/a^2$. But this ignores the fact that it is the anisotropic stress (certainly not the energy density) that balances the anisotropic curvature according to the shear-free condition. 

Let us sketch a more complete argument (leaving two steps to be proved below):
\begin{enumerate}
	\item[a)] The components of $^3\!S_{\mu\nu}$ relative to an orthonormal basis decay as $1/a^2$ in the shear-free limit (to be proved below), where $a(t)$ is the scale factor controlling distances in $\Sigma_t$,
	\item[b)] but then also anisotropic stress $\pi_{\mu\nu}$ must decay as $1/a^2$  according to the shear-free condition (\ref{shearfree:S}).
	\item[c)] The fact that the energy-momentum tensor (of the gauge field) is homogeneous quadratic in the field strength, suggests the implication (to be proved below)
	\be
	\pi_{\mu\nu}\propto 1/a^2 \;\; \Longrightarrow \;\; \rho_{\mathcal A}\propto 1/a^2 \,.
	\label{implication}
	\ee
	\item[d)] Considering (\ref{eq:evA}) with $\sigma_{\mu\nu} = 0$ we see that $\rho_{\mathcal A}\propto 1/a^2$ is only possible if $P_{\mathcal A} / \rho_{\mathcal A} = -1 / 3$,
	\item[e)] which is exactly the equation of state of a gauge field with $p\in\{0,2\}$ with spacelike field strength in the case of zero energy flux.
	\item[f)] The Maxwell case $p=1$, on the other hand, has equation of state $+1/3$ and can thus be ruled out as a candidate to balance anisotropic curvature ($^3\!S_{\mu\nu}$) on the right-hand side of (\ref{eq:shear}). 	
\end{enumerate}
To prove the statement in a), use that $a^2(t)\longmapsto a^2(t+\Delta t)$ represents a conformal transformation of three-dimensional hypersurfaces. To prove implication (\ref{implication}), start from the decomposition (\ref{ny:1})-(\ref{ny:2})  and write $\rho_{\mathcal A}$ as a function of $\pi_{\mu\nu}$. The details of these proofs are given in version 2 of this preprint \cite{ny:preprint}. This definitely excludes the 1-form as a candidate to balance anisotropic curvature, as stated in point f).

\subsection{Dynamical equivalence with FLRW \label{ch:FLRWeq}}

As noted above a comoving $p$-form gauge field with $p=2$, or equivalently $p=0$,  with spacelike field strength, has exactly the required equation of state to satisfy the shear-free condition (\ref{shearfree:S}).  In this case the gauge field has no dynamical effect on the expansion because its energy $\rho_{\mathcal A}$ and pressure $P_{\mathcal A}$ exactly cancel on the right-hand side of the Raychudhuri equation (\ref{eq:raychudhuri}). Futhermore, considering the Friedmann equation (\ref{eq:friedmann}), $\rho_{\mathcal A}$ degenerates with the three-dimensional Ricci scalar $^3\!R$ that both decay as $1/a^2$ in the shear-free limit as noted above. Hence we define an effective curvature constant
\be
K_{\rm{eff}} = \frac{a^2(t)}{6}  \cdot \left(^3\!R(t) - 2\rho_{\mathcal A}(t) \right)\,,  
\label{keff}
\ee 
which is constant in time (and space) when the congruence is shear-free. Remarkably, in the shear-free case ($\sigma_{\mu\nu}=0$), the Friedmann equation (\ref{eq:friedmann}), Raychudhuri equation (\ref{eq:raychudhuri}) and the conservation equation for perfect fluid (\ref{eq:evPF}) can be written as
\begin{align}
&H^2 + \frac{K_{\rm{eff}}}{a^2} = \frac{\rho}{3} \,,\label{FLRWequiv:1}\\
&\dot H + H^2 = -\frac{1}{6} (\rho+3P) \,,\label{FLRWequiv:2}\\
&\dot \rho + 3H(\rho+P) = 0 \,, 
\label{FLRWequiv}
\end{align}
which is exactly on the form of FLRW standard cosmologies containing only perfect fluids. We recall that 
\[
\rho = \sum_{\ell} \rho_\ell \quad \rm{and} \quad P = \sum_{\ell} P_\ell   
\]
represent the total energy and pressure of standard $\Lambda$CDM matter fields, where the index $\ell$ run over all relevant matter fields, like cold dark matter, radiation and ordinary matter. All shear-free solutions in our considered class of models can thus be mapped onto FLRW solutions, as we shall verify model by model in section \ref{ch:shearfree}. Note that this perfect screening of anisotropic curvature is a feature of the background dynamics that is expected to be broken at the level of perturbations. See \cite{pert:Pereira12} and \cite{pert:Pereira17} for perturbations in shear-free LRS Bianchi type III ($N^a_{\;\;a}=0$) models.

\section{Classification and constraints \label{ch:main:class&con}}

\subsection{Bianchi classification \label{ch:BianchiTypes}}
In this section we give a brief overview of the metric approach to Bianchi spacetimes and the classification in Bianchi types. For further details we refer to resources such as \cite{bok:MacCallum79,bok:EllisWainwright,bok:Hervik}. Conventions and the frame established here will be used subsequently in all computations, unless else is stated. 

%
We shall adopt a metric approach with line-element
\be
\mathrm ds^2 = - \mathrm dt^2 + h_{ab}(t) \mathbf W^a \mathbf W^b\,,
\ee
where Latin indices run from 1 to 3. Here the metric components $h_{ab}$ of $\Sigma_t$ depend only on the time coordinate $x^0=t$, whereas the spatial basis one-forms $\{ \mathbf W^a \}$ depend only on spatial coordinates $x^a$. Let $\{\mathbf E_a\}$ denote the corresponding basis vecors, so that  $\mathbf W^a(\mathbf E_b)=\delta^a_b$ by definition. Let $\{\mathbf Y_a\}$ denote three independent Killing fields that generate transitive isometries and form a closed algebra, or subalgebra in case there are more symmetries than spatial homogeneity. By construction $\{\mathbf Y_a\}$ span each tangent space of $\Sigma_t$. The Bianchi classification is based upon an invariant decomposition of all three-dimensional Lie algebras, whose elements are $\{\mathbf Y_a\}$.

We choose a standard left-invariant frame by choosing $\mathbf E_a=\mathbf Y_a$ at spatial origin ($x^a=0$) and Lie transporting each basis vector. 
With this convention the structure coefficients of the Lie algebra 
\be
\left[ \mathbf Y_a, \mathbf Y_b \right] = C_{ab}^{\;\;\;\; c} \mathbf Y_c 
\ee
and the commutation constants of our basis
\be
\left[ \mathbf E_a, \mathbf E_b \right] = \dd{a}{b}{c} \mathbf E_c \quad \iff \quad \bos{\mathrm d} \mathbf W^a = - \frac{1}{2} \dd{b}{c}{a} \mathbf W^b \wedge \mathbf W^c\,,
\label{def:D2}
\ee
are simply related by 
\be
D_{ab}^{\;\;\;\; c} = - C_{ab}^{\;\;\;\; c} \,.
\label{CvsD}
\ee

The classification of Bianchi spacetimes in types I-IX is based on the so-called Behr decomposition of the Lie algebra \cite{misc:BehrDecomposition}
\be
C_{ab}^{\;\;\;\;d}  = \epsilon_{abe} N^{ed} + A_e\left( \delta^e_a \delta^d_b - \delta^e_b \delta^d_a \right)
\label{decomposition} \,,
\ee  
where $\epsilon_{abc}$ is the totally anti-symmetric symbol ($\epsilon_{123}=1$) and $N^{[ab]}=0$. Since $\cc{a}{b}{c}$ transforms as a rank $(1,2)$ tensor under a constant transformation of basis, it is clear that $A_e$ transforms as a covector and that $N^{ab}$ transforms as a tensor density of weight $+1$ (because of its mixing with the anti-symmetric symbol). Nonetheless, the condition $N^a_{\;\;a}=0$ is frame invariant and defines an invariant subtype of Bianchi models that will turn out to be important for our class of shear-free models. 

Using the Jacobi identity for the three Killing vectors $\mathbf Y_a$, it follows that $A_b$ is in the Kernel of the matrix $N^{ab}$:
\be
N^{ab} A_b = 0 \,.
\label{NA0}
\ee
Hence, we can always choose a frame so that $A_b$ and $N^{ab}$ take the canonical form
\be
A_b=(A,0,0), \quad N^{ab}=\rm{diag}(N_1, N_2, N_3) \,.
\label{gauge-choice}
\ee
The basis vectors can be rescaled so that $N_1$, $N_2$ and $N_3$ take values in \{-1,0,1\}. Class A ($A=0$) and B ($A\neq 0$) models are divided into Bianchi types I-IX based on the structure of the eigenvalues of $N^{ab}$, according to table \ref{tab:Lie}. The Bianchi types form a nearly discrete set, apart from class B models of type VI$_h$ and VII$_h$, where one may introduce a quantitiy $h$ that is a scalar under transformations of bases \cite{misc:Collins72}. It controls the norm of the covector $A_a$; in the canonical basis: 
\be
A^2 = h N_2 N_3 \,.
\ee
Note that Bianchi type III is the particular case of the family VI$_h$ with $h=-1$, whereas Bianchi type VI$_0$ and VII$_0$ are the Bianchi class A special cases. An explicit basis reproducing the canonical form of $A_a$ and $N^{ab}$ in table \ref{tab:Lie} can be found in \cite{bok:MacCallum79}.

\subsection{Field strength ansatz and constraints \label{ch:constraints}}

In this section we start by stating our ansatz for the field strength, that follows from our definition \ref{def:shear-free} of a shear-free solution. Next we count the independent degrees of freedom of the field strength for the various Bianchi types.

We have already excluded $p$-form gauge fields $\boldsymbol{\mathcal A}$ with $p=1$ (Maxwell field) and $p=3$ (cosmological constant) as candidates to balance anisotropic curvature. The remaining candidates $p=0$ and $p=2$ are physically equivalent and the dynamical equations and energy-momentum tensor has been written on a unified form in section \ref{ch:0-2-form} in terms of the field strength $X_\mu$ defined in (\ref{def:eq:X}). In order to obtain a homogenous energy-momentum tensor compatible with the symmetries of Bianchi cosmologies, we shall assume $X_\mu=X_\mu(t)$. Note that, since we restrict our attention to the case of zero energy flux ($q^\mu=0$) and non-zero anisotropic stress ($\pi_{\mu\nu}\neq0$), according to our definition \ref{def:shear-free} of a shear-free solution, the temporal component of $X_\mu$ vanishes. Our ansatz for the field strength is thus
\be
X_\mu = (\; 0,X_1(t), X_2(t), X_3(t) \;) \,.
\label{ansatz:X}
\ee
Indeed, this ansatz corresponds to an inhomogeneous scalar field: $\phi(\vec x)$. Since the Lagrangian is free of a potential $V(\phi)$, this is \emph{not} incompatible with the symmetries of Bianchi cosmologies. In the 2-form description our ansatz for $X_\mu$ allows for, but is not restricted to, a homogenous gauge field $B_{\mu\nu}(t)$. Since the gauge field is observable only via the field strength, our attention will be fully devoted to $X_\mu$ below and in the forthcoming sections. 

Next we count the independent degrees of freedom of the field strength for the various Bianchi types. Specifically, we shall use equation (\ref{eq:X1}) and (\ref{eq:X2}) to determine the number of independent components of $X_a$, and their time evolution, for each Bianchi type under the ansatz (\ref{ansatz:X}). The results are summarized in table \ref{tab:Lie} and will be used in section \ref{ch:shearfree} to systematically derive all general relativistic shear-free solutions in this class of models. Similar counting in related models have been presented in \cite{main:Batakis95,main:Barrow96,ny:Normann2017}.

Equation (\ref{eq:X1}) can be written 
\be
\partial_\mu X_\nu - \partial_\nu X_\mu = -\dd{\nu}{\mu}{\gamma}  X_\gamma \,.
\ee
In our standard basis $\bos{e_0}=\partial_t$ commutes with $\mathbf E_a$ so that $\DD{0}{a}{\mu}=0$. It follows that the space-time components $(\mu,\nu)=(a,0)$ of the equation can be written
\be
\frac{\partial}{\partial t} X_a = 0 \,.
\ee
Thus $X_a$ are three constants in all Bianchi models, which implies that the energy decays as
\be \rho_{\mathcal A} = h^{ab} X_a X_b \propto a^{-2} \ee 
if $h_{ab}(t)\propto a^2(t)$, in agreement with the analysis of the shear-free limit in section \ref{ch:FLRWeq}.

In the decomposition (\ref{decomposition}) the space-space components of (\ref{eq:X1}) and (\ref{eq:X2}), respectively, can  be written as
\begin{align}
N^{ab} X_{b} + \epsilon^{abc} A_{b} X_{c}& = 0\,, \label{constraint1} \\
A_a X^a& = 0 \,. \label{constraint2}
\end{align}

\paragraph*{\textbf{Bianchi class A}} If $A=0$ equation (\ref{constraint2}) is trivial, whereas (\ref{constraint1}) gives
\begin{equation}
N_1 X_1 = 0, \quad  N_2 X_2 = 0, \quad  N_3 X_3 = 0 \,.
\label{heihei}
\end{equation}
Thus, in Bianchi class A models, $X_{a} = 0$ if $N_{a}\neq 0$ and $X_{a}$ is free if $N_{a} = 0$. The results are summarized in table \ref{tab:Lie}.  

\paragraph*{\textbf{Bianchi class B}} If $N_1=0$ and $A\neq 0$ the constraints (\ref{constraint1}) and (\ref{constraint2}) can be written
\begin{align}
N_2 X_2 - A X_3 &= 0 \,,  \\ 
N_3 X_3 + A X_2 &= 0 \,,  \\ 
X^1(t) &=  0 \,.  
\end{align}
For type III, which is type VI$_h$ with group parameter $h=-1$, this implies $X_2 = X_3$ and $X^1=0$ (note the contravariant index) or 
\be
X_{a} = X_2 \left( -\frac{h^{12}+h^{13}}{h^{11}}, \quad 1, \quad 1 \right) \,,
\ee
so we have one free parameter (denoted $X_2$). For all other class B types we obtain $X_2 = X_3 = 0$ and $X^1 = 0$. Since $\{\mathbf W^{a}\}$ is a spacelike frame by construction, $h^{11}>0$ and thus $X^1 = h^{1a} X_{a} = h^{11} X_1=0$ implies $X_1=0$. Thus $X_{a}=(0,0,0)$ for all Class B models apart from type III. The results are again summarized in table \ref{tab:Lie}.
\begin{table}
	\footnotesize
	\newcommand\T{\rule{0pt}{2.6ex}}
	\newcommand\B{\rule[-1.2ex]{0pt}{0pt}}
	\begin{center}
		\begin{tabular}{ c c c  c  c  c c c  c c c  c }
			\hline\hline 
			Class \T \B    &  Type & $N_1$ & $N_2$ & $N_3$ & $A$ & Components of $X_a$    \\ \hline
			\T\B A& I  & 0 & 0 & 0 & 0 & $(X_1, X_2, X_3)$ \\
			\B & II  & 1  & 0 & 0 & 0 & $(0, X_2, X_3)$ \\
			\B & VI$_{0}$  & 0 & 1 & $-1$ & 0 & $(X_1, 0, 0)$ \\
			\B & VII$_{0}$  & 0 & 1 & 1 & 0 & $\left(X_1, 0, 0\right)$ \\
			\B & VIII  & 1 & 1 & $-1$ & 0 & $(0,0,0)$ \\
			\B & IX  & 1 & 1 & 1 & 0 & $(0,0,0)$ \\ 
			\T\B B & III (VI$_{-1}$)  & 0 & 1 & $-1$ & 1 & $X_2\left( -\frac{h^{12}+h^{13}}{h^{11}}, 1, 1 \right)$ \\
			\B & IV  & 0 & 0 & 1 & 1 & $(0,0,0)$ \\
			\B & V  & 0 & 0 & 0 & 1 & $(0,0,0)$ \\
			\B & VI$_{h} (-1\neq h<0)$  & 0 & 1 & $-1$ & $\sqrt{-h}$ & $(0,0,0)$ \\
			\B & VII$_{h} (h>0)$  & 0 & 1 & 1 & $\sqrt{h}$ & $(0,0,0)$ \\
			\hline\hline
		\end{tabular}
		\caption{Lie algebra and independent components of the gauge-field variable $X_a$ for all Bianchi types.} 
		\label{tab:Lie}
	\end{center}
\end{table}

\section{General relativistic shear-free solutions \label{ch:shearfree}}

Building on the results above we will here systematically derive and classify shear-free solutions in all Bianchi models that can accomodate the required gauge field, as summarized in table \ref{tab:Lie}. The emphasis is on completeness, i.e. we want to find \emph{all} shear-free solutions as defined in section \ref{ch:intro}. We will thus reproduce previous solutions in the literature as well as finding new solutions. First, in subsection \ref{ch:shearfreecondition} we develop a framework, that enables us to derive and classify the most general form of the shear-free solutions, which will be employed model by model in the following subsections.

\subsection{Framework \label{ch:shearfreecondition}}

In order to find the most general form of the shear-free solutions according to the definition in section \ref{ch:intro}, we will consider a general metric tensor with off-diagonal components. This requires a strategy for avoiding false algebraic solutions; we will introduce some fundamental constraints following directly from the Cauchy-Schwarz inequalities, that will prove very useful subsequently. Our strategy for handling a multitude of gauge fields and a procedure for classifying the boundedness of the Hamiltonian will also be established.

\paragraph*{\textbf{Metric ansatz}}
We are interested in solving Einstein's equation for a most general Bianchi type geometry compatible with conformal expansion. Thus our metric ansatz will be
\be
h_{ab}(t)=a^2(t) \cdot \hat h_{ab} \,,
\label{ansatz:metric}
\ee
where
\be
\hat h_{ab} = \left(
\begin{array}{cccc}
	\vartheta^2 & \mu & \nu \\ 
	\mu & \varphi^2 & \lambda \\ 
	\nu & \lambda & \gamma^2
\end{array}\right)
\ee
is a positive definit matrix with constant components. Of course, in that case all components of the shear tensor
\be
\sigma_{ab} = \frac{1}{2} \left( \frac{\partial}{\partial t}(h_{ab}) - \frac{1}{3} h_{ab} h^{cd} \frac{\partial}{\partial t} (h_{cd})   \right)
\ee
are identically zero. For convenience we define $\det(h_{ab})=a^6(t)$ and it follows that $\hat h_{ab}$ has five independent constant components and unity determinant: 
\be
\det \hat h_{ab} \equiv \vartheta^2\varphi^2\gamma^2 + 2\mu\nu\lambda - \vartheta^2 \lambda^2-\varphi^2 \nu^2-\gamma^2 \mu^2=1 \,.
\label{deth}
\ee
The choice of frame, presented in each subsection below, is so that the basis vector commutation coefficients $D_{ab}^{\;\;\;\; c}$, upon accounting for the sign convention (\ref{CvsD}), agree with the canonical form of the Lie algebra decomposition in table \ref{tab:Lie}.

\paragraph*{\textbf{Shear-free condition}}
In the shear-free case, the shear-free condition (\ref{shear-free-con}) provides a convenient split of Einstein's equation into algebraic constraints and differential equations. Specifically, for the metric (\ref{ansatz:metric}), (\ref{shear-free-con}) reduces to purely algebraic constraints on the metric constants $\{ \vartheta^2, \varphi^2, \gamma^2, \mu, \nu, \lambda \}$. In order to derive general relativistic solutions, it is therefore very convenient to start by an analysis of the shear-free condition. The dynamics $a(t)$ is subsequently obtained from the remaining parts of Einstein's equation, which reproduces the Raychudhuri equation and the Hamiltonian constraint (Friedmann equation). This procedure is in principle straight forward. However, the constraint (\ref{deth}) does not, by itself, assure that $h_{ab}$ is positive definite, because a positive determinant allows for one positive and two negative eigenvalues. Therefore, we need a procedure to identify false algebraic solutions. 

\paragraph*{\textbf{Cauchy-Schwarz constraints}}
We need to write down constraints among the parameters $\{ \vartheta^2, \varphi^2, \gamma^2, \mu, \nu, \lambda \}$ that assure the positive definiteness of $h_{ab}$. The spatial basis vectors $\{\mathbf E_a \}$ span the tangent spaces $T_p$ of each constant time hypersurface $\Sigma_t$. Since $\Sigma_t$ is spacelike each tangent space is equipped with an ordinary inner product provided by the induced metric $h_{ab}=\mathbf E_a \cdot \mathbf E_b$. This has two implications.  First, of course, the diagonal components of $h_{ab}$ must be positive
\be
\vartheta^2>0 \,, \qquad
\varphi^2>0 \,,  \qquad 
\gamma^2>0 \,. \label{constraints}
\ee
Second, since the basis vectors are linearly independent they satisfy a strict Cauchy-Schwarz inequality
\[
(\mathbf E_{a}\cdot \mathbf E_{b})^2 < \left( \mathbf E_{a} \cdot \mathbf E_{a} \right) \left( \mathbf E_{b} \cdot \mathbf E_{b} \right), \qquad (a\neq b,\;\; \rm{no\; summation})
\nonumber
\]
which can be written
\be
\mu^2 < \vartheta^2 \phi^2 \,,  \qquad
\nu^2 < \vartheta^2 \gamma^2 \,,  \qquad 
\lambda^2 < \varphi^2 \gamma^2 \,. \label{CSconstraints}
\ee
We will refer to these three inequalities as the \emph{Cauchy-Schwarz constraints}. Together with (\ref{constraints}) they will be employed to eliminate false algebraic solutions of (\ref{shear-free-con}) in the following subsections. They will also be used to fix the sign of $\rho_{\mathcal{A}}$ unambiguously.
In appendix \ref{ch:sylvester} we prove, using Sylvester's criterions, that these constraints together with (\ref{deth}) assure the positive definiteness of $h_{ab}$. It is clear from that proof that the six constraints are not independent; nevertheless, each of them will prove useful for eliminating false solutions.

\paragraph*{\textbf{Boundedness of the Hamiltonian}} 
In this work we do not assume any energy condition from the start. Our goal is to find all shear-free solutions, followed by a classification of the Hamiltonian boundedness. In section \ref{ch:Hamiltonian} we established an equivalence between the weak energy condition and the boundedness of the Hamiltonian for the considered family of gauge theories. This implies that if the gauge field energy $\rho_{\mathcal A}$ seen by a comoving observer is negative, the Hamiltonian is not bounded from below. This situation is only possible in case we replace the prefactor $-1/2$ in the action (\ref{action}) by some positive number. If considering a free parameter $\alpha$ in the action (\ref{action}), i.e. $S\rightarrow \alpha S$, a necessary and sufficient condition for the Hamiltonian to be bounded from below is $\rm{sign}(\alpha)>0$. Instead of introducing a free parameter, we shall keep the action on the canonical form (\ref{action}) throughout, but consider both real and (purely) imaginary values for the field strength $\boldsymbol{\mathcal F}$. Since $S$ is quadratic in $\boldsymbol{\mathcal F}$, imaginary values simply correspond to flipping sign $S \rightarrow -S$. In the following subsections this observation will be used to classify exact solutions with respect to energy conditions. First we need to generalize this procedure to the case of multiple gauge fields.

\paragraph*{\textbf{Multiple gauge fields}} We consider a collection of $n$ independent $p\in\{0,2\}$-form gauge fields $\bos{\mathcal A}^{(l)}$, where the index $\ell$ label each gauge field and run from 1 to $n$. The total energy-momentum tensor is given by (\ref{totalEMT}). Let $X_a^{(\ell)}$ denote the associated field strength variable, as defined in (\ref{def:eq:X}). Considering the field strength constraints summarized in table \ref{tab:Lie}, we note that among the spacetimes that can accommodate a gauge field, several admit only one independent compontent of $X_a$. This is the case for Bianchi type VI$_0$, VII$_0$ and III. When we consider multiple gauge fields, the constraints therefore imply that the field strengths are aligned and span a line. Hence the generalization from $n=1$ to arbitrary $n$ is rather trivial in these cases. Let us introduce some notation, that will also be employed in the technically more interesting case of Bianchi type II. 

We denote the independent component of each gauge field by $\Psi_\ell$. For instance in Bianchi type VI$_0$ and VII$_0$ the field strength can then be written $X_\mu^{(\ell)} = (0,\Psi_\ell,0,0)$. Note that, for a given metric, the total energy-momentum tensor of the gauge field sector is controlled uniquely by the tensor 
\be
Y_{\mu\nu}\equiv \sum_{\ell=1}^{n} X_\mu^{(\ell)} X_\nu^{(\ell)} \,.
\ee
The total anisotropic stress can then be specified as
\be
\pi_{ab} =  Y_{ab} - \frac{1}{3} h_{ab}\, h^{ij} Y_{ij} \,,
\label{genialt}
\ee
For instance, in the case Bianchi of type VI$_0$ and VII$_0$ the non-vanishing components of $Y_{\mu\nu}$ are
\be 
Y_{ab} = \rm{diag}\left( \Psi^2,\; 0,\; 0 \right), \quad \Psi^2\equiv\sum_{\ell=1}^n \Psi_\ell^2 \,.
\ee
In all Bianchi models except type I and II, the $n$ physical field strengths $X_a^{(\ell)}$ can always be mapped onto a single abstract field strength $Y_a$ which produces the same tensor $Y_{ab}$ and hence identical energy-momentum tensor. We shall refer to $Y_a$ as the \emph{generator} of the energy-momentum tensor. 

\begin{definition}[generator]
	If the total energy-momentum tensor associated with the collection of $n$ gauge fields can be reproduced by a single abstract gauge field with field strength $Y_a$, then we refer to $Y_a$ as a generator of the energy-momentum tensor.   
	\label{def:generator}
\end{definition}

Note that $Y_a$ is \emph{not} the total field strength $\sum_\ell X_a^{(\ell)}$; the generator is a purely abstract gauge field with a field strength $Y_a$ that, by definition, happens to produce the same total energy-momentum tensor as the $n$ independent physical gauge fields. For instance in Bianchi type VI$_0$ and VII$_0$ the generator is $Y_a = (\Psi,0,0)$, which reproduces the tensor $Y_{ab}$ above.


The case of Bianchi type II, in which the physical field strengths span a plane, rather than a line, is much more interesting at the technical level. Using an automorphism we will be able to map the $2n$ dimensional configuration space onto two independent generators $Y_a^{(1)}$ and $Y_a^{(2)}$, each with a single independent component (say, $\Psi$ and $\Theta$). This generalizes the concept of a generator as defined above to the smallest set of abstract gauge fields that reproduce the total energy-momentum tensor. The details are saved for section \ref{ch:BII}.

Below the shear-free condition will be analyzed based on generators instead of the physical gauge fields. The existence of a shear-free solution where all physical fields have a Hamiltonian bounded from below, requires that the generators are real: $\Theta,\Psi \in \mathbb R$.


\subsection{Bianchi type II \label{ch:BII}}

In section \ref{ch:constraints} we found that Bianchi type II is the only Bianchi spacetime with a non-Abelian Lie algebra in which the field strength has more than one independent component. Specifically the field strength covectors $X_a^{(\ell)}$ span a plane, which means that when we consider a multitude of gauge fields ($n>1$), the field strengths are generally not aligned. Below we will identify an automorphism group that allows us to deal with this situation, without any loss of generality. Automorphisms can be viewed as gauge-transformations associated with the freedom in choice of frame and are well-recognized for their importance in identifying the true physical degrees of freedom in Bianchi cosmologies \cite{bok:EllisWainwright}. Automorphisms in this context are transformations of bases, itself with a group structure, that preserve the Lie algebra. Different types of automorphisms have been identified and presented in the literature \cite{auto:Jantzen01}, \cite{auto:Christodoulakis00}. Here we will identify a special class of automorphisms associated with diffemorphisms that leave not only the Lie algebra invariant, but indeed also the form of the basis objects that defines the frame. 
As we shall see this enables us to map $n$ arbitrary gauge fields onto two abstract gauge fields, the generators of the energy-momentum tensor, whose field strength components $Y^{(1)}_a$ and $Y^{(2)}_a$ have a particularly simple form. It is then relatively straight forward to carry out a fully general analysis of the shear-free condition for $n$ arbitrary gauge fields. In fact, we shall see that the shear-free condition can be rephrased in geometrical statements about the two generators.  

\subsubsection{Choice of frame, automorphisms and symmetries \label{ch:BII:frame}}
\paragraph*{\textbf{Frame}}
We use coordinates in which the basis dual vectors and vectors are on the form:\footnote{Another popular choice is $\mathbf V^a= \left(\bos{\mathrm d} x+ y\bos{\mathrm d} z, \bos{\mathrm d} y, \bos{\mathrm d} z \right)$, which has identical Lie algebra. Consequently, the metrics $h_{ab}\mathbf V^a\mathbf V^b$ and $h_{ab}\mathbf W^a\mathbf W^b$ have identical Riemann tensors $R_{cdef}$ for all $h_{ab}$. We use the less compact form here in order to clearly discuss symmetries. Note that the bases $\mathbf W^a$ and $\mathbf V^a$ are the same fields expressed in different coordinates related by  $(x,y,z)\longmapsto (x-yz/2, y, z)$.} 
\begin{align}
\mathbf W^a &= \left( \bos{\mathrm d}x - \frac{1}{2} \left(z \bos{\mathrm d}y - y \bos{\mathrm d}z\right), \quad   
\bos{\mathrm d}y, 
\quad
\bos{\mathrm d}z \right) \,, \label{BII:basis:1} \\	
\mathbf E_a &= \left( \frac{\partial}{\partial x}, \quad   
\frac{z}{2} \frac{\partial}{\partial x} + \frac{\partial}{\partial y}, \quad
-\frac{y}{2}\frac{\partial}{\partial x} + \frac{\partial}{\partial z} \right) \,.
\label{BII:basis:2}
\end{align}

\paragraph*{\textbf{Automorphisms}}
We will now identify a transformation of frame that preserves the form of all basis objects (\ref{BII:basis:1})-(\ref{BII:basis:2}). The transformation acts on our anholonomic basis in the same way as a standard coordinate transformation acts on a holonomic basis (coordinate basis). This automorphism will then be applied in the following subsections to study the shear-free condition in Bianchi type II with no loss of generality. This point is cruical for the proof of Theorem \ref{int:theorem2} and further details are given in version 2 of this preprint \cite{ny:preprint}.


Motivated by the structure of the basis objects (\ref{BII:basis:1})-(\ref{BII:basis:2}), we consider a coordinate transformation that rotates with respect to the $x$-axis
\be
x^a \longmapsto x^{\tilde a} = S^{\tilde a}_{\;\;b} x^b \,, \quad S^{\tilde a}_{\;\;b} \equiv ( \mathcal S)_{ab}\,, \quad \left( S^{a}_{\;\;\tilde b} \equiv ( \mathcal S^T)_{ab} \right) \,,
\label{BII:coordinatetransformation}
\ee
where $\mathcal S$ is the matrix 
\be 
\mathcal S = \left(\begin{array}{ccc}
	1 & 0 & 0  \\ 
	0 & \cos \alpha & -\sin \alpha \\
	0 & \sin \alpha & \cos \alpha  
\end{array}\right) 
\ee
and $\alpha$ is an arbitrary constant. We will label individual coordinates as $x^a=(x,y,z)$ and $x^{\tilde a}=(\tilde x,\tilde y,\tilde z)$. Notice that the Jacobian $\partial x^{\tilde a} / \partial x^b$ here is $S^{\tilde a}_{\;\;b}$ because the coordinate transformation is linear. As in the case of holonomic bases we redefine the basis using the Jacobian:
\be
\mathbf W^a \longmapsto S^{\tilde a}_{\;\;b} \mathbf W^b = \mathbf W^{\tilde a}, \qquad \mathbf E_a \longmapsto S^{b}_{\;\;\tilde a} \mathbf E_b = \mathbf E_{\tilde a} \,,
\label{BII:symmetrytransformation}
\ee
where
\begin{align}
\mathbf W^{\tilde a} &= \left( \bos{\mathrm d}\tilde x - \frac{1}{2} \left(\tilde z \bos{\mathrm d} \tilde y - \tilde y \bos{\mathrm d}\tilde z\right), \quad   
\bos{\mathrm d}\tilde y, 
\quad
\bos{\mathrm d}\tilde z \right) \,, \\	
\mathbf E_{\tilde a} &= \left( \frac{\partial}{\partial \tilde x}, \quad   
\frac{\tilde z}{2} \frac{\partial}{\partial \tilde x} + \frac{\partial}{\partial \tilde y}, \quad
-\frac{\tilde y}{2}\frac{\partial}{\partial \tilde x} + \frac{\partial}{\partial \tilde z} \right) \,.
\end{align}
Notice that the form of the original basis (\ref{BII:basis:1})-(\ref{BII:basis:2}) is preserved, expressed in the new coordinates. It follows directly that also the Lie algebra is preserved, so the transformation is an automorphism in the usual sense \cite{bok:EllisWainwright}. Clearly, the set of all automorphisms considered here has a $SO(2)$ group structure. It is only a subgroup of the full set of basis transformations that preserves the type II Lie algebra, but will prove general enough for our purposes. It should be mentioned, perhaps, that the generalization of our fixed axis rotation to three-dimensional rotations does not preserve the form of the basis, i.e. is not an automorphism.  The higly non-trivial step of the symmetry transformation (\ref{BII:symmetrytransformation}) is the second step, i.e. the six equations $S^{\tilde a}_{\;\;b} \mathbf W^b = \mathbf W^{\tilde a}$ and $S^{b}_{\;\;\tilde a} \mathbf E_b = \mathbf E_{\tilde a}$. 


Thus we have established a class of transformations, defined by (\ref{BII:coordinatetransformation}) and (\ref{BII:symmetrytransformation}), that is consistent with our choice of frame (\ref{BII:basis:1})-(\ref{BII:basis:2}). We could have used this gauge freedom to simplify the form of the metric components $h_{ab}$. Instead we shall leave the metric on the general non-diagonal form (\ref{ansatz:metric}) and fix the gauge by simplifying the structure of the total energy-momentum tensor. In practice we apply the automorphism (\ref{BII:symmetrytransformation}) by applying the standard transformation law for tensors.

\subsubsection{A single gauge field ($n=1$) \label{ch:BII:n1}}
A most general gauge field compatible with the constraints worked out in section \ref{ch:constraints} is
\be 
X_a = (0, \Theta, \Psi)\,, 
\ee 
where $\Psi$ and $\Theta$ are arbitrary constants. Employing the automorphism (\ref{BII:symmetrytransformation}), we note that $X_a$ can be written on the simpler form 
\be X_a = (0,\Theta, 0) \,, \ee
with no loss of generality. This gauge choice corresponds to choosing a frame in which $\mathbf W^2$ is aligned with the field strength. The components of the anisotropic stress $\pi^a_{\;\;b}$
and the electric Weyl tensor $E^{a}_{\;\;b}$ are written down in version 2 of this preprint \cite{ny:preprint}. From them it follows that the shear-free condition $\pi^a_{\;\;b} = 2 E^a_{\;\;b}$ cannot be satisfied. 

The absence of a shear-free solution (realized by a single gauge field) is perhaps surprising given that the Bianchi type II metric has a particular case of LRS symmetry.  However, the LRS axis is not aligned with the field strength $X_a$. As the symmetry axes of spatial geometry and matter are not aligned, they cannot be brought in balance.


\subsubsection{A collection of gauge fields ($n>1$) \label{ch:BII:n2}}

As mentioned above, the Bianchi type II is special in the sense that it is the only model with non-zero spatial curvature in which the gauge field $X_a$ has two degrees of freedom. With a single gauge field this did not lead to any technical difficulty since one of the components could be rotated away using the gauge freedom in choice of frame. This changes a bit when we consider a collection of $n$ gauge fields.

We introduce a collection of $n$ independent gauge fields with field strengths
\be X^{(\ell)}_a = (0,\Theta_\ell,\Psi_\ell) \,. \ee 
Again, $\Theta_\ell$ and $\Psi_\ell$ are arbitrary constants, so the gauge field sector is associated with a $2n$ dimensional configuration space. However, in order to derive general relativistic solutions the relevant object is the \emph{total} anisotropic stress which, according to (\ref{genialt}), is controlled by the matrix
\be
Y_{ab}\equiv \sum_{\ell=1}^{n} X_a^{(\ell)} X_b^{(\ell)} = 
\sum_{\ell=1}^{n}\left(
\begin{array}{ccc}
	0 & 0 & 0  \\ 
	0 & \Theta_\ell^2 & \Theta_\ell \Psi_\ell  \\
	0 & \Theta_\ell \Psi_\ell & \Psi_\ell^2 
\end{array}\right) \,.
\ee
Note that the symmetric tensor $Y_{ab}$ has only 3 independent components. Also note that $Y_{ab}$ is a sum of positive semidefinite matrices with eigenvalues $\{0, 0, \Theta_\ell^2+\Psi_\ell^2\}$. This implies, from definition, that also $Y_{ab}$ is positive semidefinite. Here we have assumed that $\Theta_\ell$ and $\Psi_\ell$ are real, which corresponds to boundedness of the Hamiltonian from below. Finally, note that the matrix $Y_{ab}$ happens to be on the form that can be diagonalized using the automorphism defined by transformation (\ref{BII:symmetrytransformation}). We therefore choose a basis in which the matrix is on the form
\be
Y_{ab} = \rm{diag}(0, \Theta^2, \Psi^2) \,,
\label{finalform}
\ee
where $\Theta^2,\Psi^2 \ge 0$ since $Y_{ab}$ is positive semidefinite. To summarize, given a collection of $n$ independent gauge fields with arbitrary field strengths, the anisotropic stress $\pi_{ab}$ is controlled by two independent numbers, $\Theta^2$ and $\Psi^2$. 
The important point is that $\pi_{ab}$ is blind to the full $2n$ dimensional configuration space, but only see the matrix $Y_{ab}$, whose spectrum is two-dimensional.

Note that the matrix (\ref{finalform}) can always be represented by a pair of abstract gauge fields, the generators of the total energy-momentum tensor: 
\be
Y_a^{(1)} = (0, \Theta, 0), \quad Y_a^{(2)} = (0, 0, \Psi) \,. 
\label{BII:ansatz}
\ee 
Curiously, finding general relativistic solutions with $n$ arbitrary gauge fields is equivalent to analyzing the particular case $n=2$ under the ``field strength ansatz'' (\ref{BII:ansatz}). In fact, the constraints derived from the shear-free condition (\ref{shear-free-con}) below can be rephrased in geometrical statements about the two generators $Y^{(1)}_a$ and $Y^{(2)}_a$. Namely, (\ref{dia1}) and (\ref{dia2}) imply that they have the same norm ($h^{{a}{b}} Y_{a}^{(1)} Y_{b}^{(1)} = h^{{a}{b}} Y_{a}^{(2)} Y_{b}^{(2)}$), whereas (\ref{offdia1}) implies that they are perpendicular ($h^{{a}{b}} Y_{a}^{(1)} Y_{b}^{(2)} = 0$).


\paragraph*{\textbf{Shear-free condition}}

A careful analysis shows that $\pi^a_{\;\;b}=2E^a_{\;\;b}$ can be reduced to three independent constraints:
\begin{align} 
\vartheta^4  &= -\Theta^2 (\vartheta^2\gamma^2-\nu^2) \,, \label{dia1} \\
\vartheta^4  &= -\Psi^2 (\vartheta^2\varphi^2-\mu^2) \,, \label{dia2} \\
\mu\nu &= \lambda\vartheta^2 \,. \label{offdia1} 
\end{align}
First, notice that both fields must be present ($\Theta^2\neq0$, $\Psi^2\neq0$) to have a consistent metric ($\vartheta^2$ is a diagonal component). Next, using the Cauchy-Schwarz constraints (\ref{CSconstraints}), 
it follows that $\Theta^2$ and $\Psi^2$ are negative. But then (\ref{dia1})-(\ref{dia2}) imply upper bounds $\Theta^2\le -\vartheta^2/\gamma^2$ and $\Psi^2\le -\vartheta^2/\varphi^2$ since we must have $\mu^2\ge0$ and $\nu^2\ge0$ for the off-diagonal metric components $\mu$ and $\nu$. The negative signs of $\Theta^2$ and $\Psi^2$ imply that any formal shear-free solution requires at least one gauge field with a Hamiltonian that is not bounded from below. Notice that the negative signs are consistent with our convention (\ref{deth}) that can then be written $\vartheta^6 = \Theta^2 \Psi^2$.


Combined with the Cauchy-Schwarz the constraints above imply a matter sector with a negative total energy density
\be
\rho_{\mathcal A} = -\frac{\vartheta^4}{a^2} \,, 
\ee
but allow for consistent three-dimensional geometries with negative curvature scalar\footnote{An example on a metric and gauge field consistent with (\ref{dia1})-(\ref{offdia1}) is given by $\vartheta^2=\mu=\nu=\lambda=-\Theta^2=-\Psi^2=1$ and $\varphi^2=\gamma^2=2$. In that case $\hat h_{ab}$ has three positive eigenvalues $\{1, 2\pm\sqrt{3} \}$ and unity determinant.}
\be
^3\!R = -\frac{\vartheta^4}{2a^2} \,.
\ee

\paragraph*{\textbf{General relativistic solution}} When the algebraic constraints summarized above are satisfied, Einstein's equation with the energy-momentum tensor (\ref{totalEMT}) can be written:
\begin{align}
H^2 + \frac{\vartheta^4}{4a^2} &= \frac{1}{3} \rho \,,  \\
\dot H + H^2 &= -\frac{1}{6} (\rho+3P) \,, 
\end{align}
which is the Friedmann and Raychaudhuri equations. Comparing with (\ref{FLRWequiv:1}) we see that the shear-free solution is dynamically equivalent to a closed FLRW model with a positive effective curvature constant $K_{\rm{eff}}=\vartheta^4/4$. Since it requires a negative total gauge field energy density $\rho_\mathcal{A}$, the Hamiltonian is not bounded from below. Note that the solution is not restricted to the LRS subtype. Its main properties are summarized in table \ref{tab:shearfree}.


\subsection{Bianchi type VI$_0$ \label{ch:VI}}
Here we find a new shear-free solution which belongs to the subtype $N^a_{\;\;a}=0$ of Bianchi type VI$_0$. Curiously, it has negative spatial curvature, but expands like a flat FLRW model.

\paragraph*{\textbf{Frame}} For all Bianchi type VI$_h$ models we will use the basis: 
\be
\mathbf W^1 =  \bos{\mathrm d}x ,    
\mathbf W^2 = e^{Ax} \left(\cosh x \; \bos{\mathrm d}y - \sinh x \; \bos{\mathrm d}z\right), 
\mathbf W^3 = e^{Ax} \left(-\sinh x \; \bos{\mathrm d}y + \cosh x \; \bos{\mathrm d}z\right) 	\,,
\label{BVI:basis}
\ee
where $A\equiv \sqrt{-h}=\sqrt{|h|}$. 
Here we will consider the group parameter $h=A=0$, whereas in section \ref{ch:BIII} we will use the same basis to study Bianchi type III, which is the particular case of a VI$_h$ model with group parameter $h=-A=-1$.

\paragraph*{\textbf{Shear-free condition}}
In type VI$_0$ the generator of the energy-momentum tensor is
\be
Y_{a}=(\Psi,0,0) \,,
\label{J:VI0}
\ee 
where $\Psi$ is a constant. There is only one solution of the shear-free condition $\pi^{a}_{\;\;b} = 2E^{a}_{\;\;b}$
that satisfies the Cauchy-Schwarz constraints:
\be
\varphi^2 = \gamma^2, \quad \Psi^2=  -2 \,.
\ee
Since $N^{ab}=\rm{diag}(0, 1, -1)$, the metric constraint $\phi^2=\gamma^2$ implies that $N^{ab}$ is traceless:
\be
N^{a}_{\;\;a} = h_{ab} N^{ab} = h_{22} - h_{33} = a^2(t)(\varphi^2-\gamma^2) = 0 \,.
\ee 
Thus the shear-free type VI$_0$ solution belongs to the invariant subtype $N^a_{\;\;a}=0$. 

When the shear-free condition holds, the total energy density of the gauge fields can be written
\be
\rho_{\mathcal A} = -\frac{\varphi^4-\lambda^2}{a^2}
\ee
and the three-dimensional Ricci scalar  
\be
^3\!R = -\frac{2(\varphi^4-\lambda^2)}{a^2} \,.
\ee
Notice that to have a valid 3-geometry, the Cauchy-Schwarz constraint (\ref{CSconstraints}) requires $\varphi^4-\lambda^2 > 0$, which implies $\rho_{\mathcal A}<0$ and $^3\!R<0$.  

\paragraph*{\textbf{General relativistic solution}} 
For the valid metric Einstein's equation, with energy-momentum tensor given by (\ref{totalEMT}), can be written as 
\begin{align}
&H^2 = \frac{1}{3} \rho \,,  \\
\dot H + &H^2 = -\frac{1}{6} (\rho+3p) \,.
\end{align}
which are the Friedmann and Raychudhuri equations, respectively. Comparing with (\ref{FLRWequiv:1}), this corresponds to a vanishing effective curvature constant $K_{\rm{eff}}=0$. Hence we have a shear-free solution with anisotropic curvature ($^3\!S_{ab}\neq0$) that is dynamically equivalent to a flat FLRW model. Since the shear-free condition requires a negative total gauge field energy density $\rho_\mathcal{A}$, the Hamiltonian is not bounded from below. 
The main properties are summarized in table \ref{tab:shearfree}.


Our result fits nicely into an old result of perfect fluid Bianchi models, in which the subtype $N^a_{\;\;a}=0$ of Bianchi type VI$_0$ is the unique non-LRS class A spacetime that admits an expansion symmetry with respect to some fixed axis \cite{main:Ellis1969}. With the introduction of gauge field(s) we have shown that this expansion symmetry can be promoted to three dimensions, i.e. conformal expansion.

\subsection{Bianchi type VII$_0$}
An analysis like those above shows that there is no shear-free solution in Bianchi type VII$_0$ that belongs to our considered class.  Specifically, the shear-free condition enforces LRS symmetry, but since the LRS Bianchi type VII$_0$ is similar to LRS Bianchi type I, the shear-free solution is trivial, i.e. spatially flat, and falls outside our definition \ref{def:shear-free}. Details are given in version 2 of this preprint \cite{ny:preprint}.

\subsection{Bianchi type III \label{ch:BIII}}
Here we show that the most general shear-free solution in Bianchi type III in fact belongs to the subtype $N^a_{\;\;a}=0$, which is a LRS model with spatial sections $\mathcal R \times \mathcal H_2$. The shear-free solution obtained here, realized by a $p\in\{0,2\}$-form gauge field, is therefore not genuinly new, but a unification of the LRS Bianchi type III solutions discovered in \cite{main:Carneiro01} with a scalar field ($p=0$) and in \cite{main:Koivisto11} with a 2-form gauge field.   

\paragraph*{\textbf{Frame}} Bianchi type III is the particular case of the Bianchi type VI$_h$ family with group parameter $h=-1$. Accordingly, we use the basis (\ref{BVI:basis}) with $A=-h=1$. 

\paragraph*{\textbf{Shear-free condition}}

The generator of the energy-momentum tensor in type III is
\be
Y_a = \left( -\frac{h^{12}+h^{13}}{h^{11}} \Psi, \quad \Psi, \quad \Psi \right) \,,
\ee
where $\Psi$ is a constant. A careful analysis of the shear-free condition $\pi^{a}_{\;\;{b}}=2E^{a}_{\;\;b}$ gives two independent constraints, which are the diagonal components $(2,2)$ and $(3,3)$. The former is 
\be 
\frac{\Psi^2(2\gamma^2-\phi^2-\lambda)}{\phi^2\gamma^2-\lambda^2} = (2\phi^2-\gamma^2+\lambda)(\phi^2+\gamma^2-2\lambda) \,,
\ee
whereas the latter is the same equation up to $\phi^2 \leftrightarrow \gamma^2$. There are two algebraic solutions of these equations. The first one is $\Psi=0$ and $2\lambda=\phi^2+\gamma^2$. This is a trivial solution with vanishing $\pi^{a}_{\;\;b}$ and $E^{a}_{\;\;b}$. The Cauchy-Schwarz constraint $\lambda^2<\phi^2\gamma^2$ can then be written $(\phi^2-\gamma^2)^2<0$ so it does not represent a valid 3-geometry. The interesting solution is 
\be
\varphi^2=\gamma^2, \quad \Psi^2=2(\varphi^2+\lambda)^2(\varphi^2-\lambda) \,.
\label{b3:shearfree}
\ee
In this case the Cauchy-Schwarz constraint gives $\phi^4>\lambda^2$, which implies a positive total energy density in the gauge field sector
\be
\rho_{\mathcal A} = \frac{1}{2} Y_{a} Y^{a} = \frac{2(\varphi^4-\lambda^2)}{a^2(t)}
\ee
and a negative three-dimensional Ricci scalar
\be
^3\!R = -\frac{8(\varphi^4-\lambda^2)}{a^2(t)} \,.
\ee
We note that the metric constraint $\phi^2=\gamma^2$ is equivalent to $N^{ab}$ being traceless:
\be
N^{a}_{\;\;a} = h_{ab} N^{ab} = h_{22} - h_{33} = a^2(t)(\varphi^2-\gamma^2) = 0 \,.
\ee
Thus the shear-free type III solution belongs to the invariant subtype $N^a_{\;\; a} = 0$.

\paragraph*{\textbf{General relativistic solution}}
Using (\ref{b3:shearfree}) the Einstein equation, with energy-momentum tensor \ref{totalEMT}, can be summarized by
\begin{align}
H^2 - \frac{2(\varphi^4-\lambda^2)}{a^2(t)} = \frac{\rho}{3} \,,  \\
\dot H + H^2 = -\frac{1}{6} (\rho+3p) \,,
\end{align}
which are the Friedmann and Raychudhuri equations, respectively. This corresponds to a negative effective curvature constant $K_{\rm{eff}}=-2(\varphi^4-\lambda^2)$, and hence the shear-free solution is dynamically equivalent to an open FLRW model. Since a positive total energy density $\rho_\mathcal{A}$ is required, it has a Hamiltonian bounded from below, which is unique among the shear-free solutions within the considered class of models. The main properties are summarized in table \ref{tab:shearfree}.

An important question is if the shear-free solution derived above can be identified with the Kantowski-Sachs type metric, i.e. (\ref{b3ks:shear-free-metric}) with $k<0$ \cite{bok:KantowskiSachs66,bok:Kantowski66}. We have shown that the most general Bianchi type III shear-free solution belongs to the subtype $N^a_{\;\;a}=0$.  It is known that if the group is type III with $N^a_{\;\;a}=0$ the space is necessarily LRS \cite{main:Ellis1969}. In the subsection below we explicitly identify a coordinate transformation that confirms that the spatial sections are of type $\mathcal R \times \mathcal H_2$, i.e. the product between a flat direction and the maximally symmetric 2-space of negative curvature. To conclude, the shear-free solution derived above is a LRS solution, with field strength $X_a$ aligned with the LRS axis. This solution necessarily intersects with the shear-free solutions found in \cite{main:Carneiro01} and \cite{main:Koivisto11}; see the review in section 6 of the preprint \cite{ny:preprint} (version 2) for further comments.

\subsubsection{Bianchi type III with trace-free $N^{ab}$ has spatial sections $\mathcal R \times \mathcal H_2$}
\label{ch:BIIIiso}

The frame $\{\mathbf W^i\}$ used in the calculations above is given by (\ref{BVI:basis}) with $h=-A=-1$. It gives the canonical form of the Lie algebra given in table \ref{tab:Lie}. We introduce a rotated basis
\be
\tilde{\mathbf W}^i = \{  \mathbf W^1,\; \mathbf W^2+\mathbf W^3,\; -\mathbf W^2+\mathbf W^3 \} = \{  \mathrm d \tilde x,\; \mathrm d \tilde y,\; e^{2\tilde x} \mathrm d \tilde z \} \,,
\ee
where $(\tilde x, \tilde y, \tilde z) = (x, y+z, -y+z)$. In the new basis $A_{d}=(1,0,0)$ is unchanged whereas the $N^{ab}$ becomes off-diagonal with non-zero components $\tilde N^{23} = \tilde N^{32} = -1$.\footnote{$A_a$ is a covector and transforms accordingly, whereas the symmetric tensor density transforms as $\tilde N^{ab}=(1/\det M) M^{a}_{\;\;c} M^{b}_{\;\;d} N^{cd}$. Here the transformation matrix is defined by $\tilde{\mathbf{W}}^{a} = M^{a}_{\;\;b} \mathbf W^b$.} Hence the trace-free condition
\be
0=N^{a}_{\;\;a} =\tilde N^{a}_{\;\;a} = \tilde h_{ab} \tilde N^{ab} = -2\tilde h_{32} = -2\tilde \lambda 
\ee
implies $\tilde \lambda=0$. It follows that the most general three-dimensional type III geometry with trace-free $N^{ab}$ can be represented by the metric
\be
d\Sigma^2 = \tilde h_{ab} \tilde W^{a} \tilde W^{b} = \tilde \vartheta^2 \mathrm d \tilde x^2 + \tilde \varphi^2 \mathrm d \tilde y^2 + \tilde \gamma^2 e^{4\tilde x} \mathrm d \tilde z^2 + 2\tilde \mu \mathrm d \tilde x \mathrm d \tilde y + 2\tilde \nu e^{2\tilde x}\mathrm d \tilde x \mathrm d \tilde z \,.
\ee 
As a consequence of $\tilde \lambda =0$, this metric turns out to be diagonalizable. Upon the coordinate transformation
\be 
x = \tilde x \,, \quad y = \tilde y+\frac{\tilde \mu}{\tilde \varphi^2} \tilde x \,, \quad z = \tilde z - \frac{\tilde \nu}{2\tilde \gamma^2} e^{-2\tilde x} \,, 
\ee
the line element takes the form 
\be
d\Sigma^2 =  \vartheta^2 \mathrm d x^2 + \varphi^2 \mathrm d  y^2 +  \gamma^2 e^{4 x} \mathrm d z^2 \,,
\label{BIII:diagonal}
\ee
where we have defined $\vartheta^2 = \tilde \vartheta^2 - \frac{\tilde \mu^2}{\tilde\varphi^2} - \frac{\tilde \nu^2}{\tilde\gamma^2}$, $\varphi^2 = \tilde \varphi^2$ and $\gamma^2 = \tilde \gamma^2$.\setcounter{footnote}{0}\footnote{The new coordinates $(x,y,z)$ and metric coefficients $(\vartheta, \varphi, \gamma)$ must not be confused by the original ones used in section \ref{ch:BIII}.} 
Thus, up to diffeomorphisms, we have shown that the diagonal metric (\ref{BIII:diagonal}) represents the most general Bianchi type III geometry with trace-free $N^{ab}$. It is easy to verify from the algebra of the Killing fields that the geometry is $\mathcal R \times \mathcal H_2$, see version 2 of this preprint \cite{ny:preprint} for details. Thus we recognize (\ref{BIII:diagonal}) as the LRS Bianchi type III metric.

\subsection{Kantowski-Sachs metric \label{ch:review}}
Finally we briefly comment on the case of the Kantowski-Sachs metric
\be
\mathrm ds^2 = -dt^2 + a^2(t) \left( dx^2 +  dy^2 + k^{-1} \sin^2{(k^\frac{1}{2}y)}dz^2 \right),\quad k>0 \,,
\label{b3ks:shear-free-metric}
\ee
which is the unique case of a spatially homogeneous spacetime that falls outside the Bianchi classification. It is easy to verify the existence of a shear-free solution with $K_{\rm{eff}}=k/2$ which is related to the LRS Bianchi type III solution (section \ref{ch:BIII}) by $^3\! R \rightarrow -^3\! R$ and $\rho_{\mathcal A} \rightarrow -\rho_{\mathcal A}$ and therefore, like the Bianchi type II and VI$_0$ shear-free solutions, does not have a gauge field Hamiltonian bounded from below. This result agrees with Koivisto et.al \cite{main:Koivisto11} and concludes our investigation of shear-free solutions in the class of models given by definition \ref{def:shear-free}.

\paragraph{Acknowledgments}
We thank Sigbj\o rn Hervik, Tore A. Kro, Thiago S. Pereira and Hans A. Winther for very enlightening discussions and comments on the manuscript.

\newpage
\appendix
\section{Cauchy-Schwarz inequalities from Sylvester's criterions \label{ch:sylvester}} 
Here we shall derive the six constraints (\ref{constraints}\,:abc)-(\ref{CSconstraints}\,:abc) directly from the requirement that $h_{ab}v^av^b>0$ for all vectors $v^a\in T_p$. We remind that $h_{ab}$ is positive definite iff the leading principal minors of $h_{ab}$ are positive (known as Sylvester's criterions): 
\begin{align}
\vartheta^2 &> 0  \label{sylvester:a} \,, \\
\vartheta^2 \varphi^2 - \mu^2 &> 0 \label{sylvester:b} \,, \\
\gamma^2(\vartheta^2\varphi^2-\mu^2) + 2\lambda\mu\nu - \vartheta^2\lambda^2-\varphi^2\nu^2 &> 0 \,. \label{sylvester:c}
\end{align}
Note that (\ref{sylvester:c}) is assured by our convention (\ref{deth}). Furthermore, (\ref{sylvester:a}) and (\ref{sylvester:b}) amount to constraints (\ref{constraints}\,:a) and (\ref{CSconstraints}\,:a), respectively. It is clear that given our convention (\ref{deth}), the 6 constraints (\ref{constraints})-(\ref{CSconstraints}) are not independent. In order to show how they are related, it remains to derive (\ref{constraints}\,:bc) and (\ref{CSconstraints}\,:bc) from Sylvester's criterions.

It is easy to see that (\ref{constraints}\,:b) follows from (\ref{sylvester:a}) and (\ref{sylvester:b}). Next (\ref{constraints}\,:c) can be proved by contradiction by assuming $\gamma^2=0$ in (\ref{sylvester:c}):    
\be  
0  >  \vartheta^2\lambda^2 - 2\lambda\mu\nu + \varphi^2\nu^2 \,. 
\ee
Notice that the right-hand side is a quadratic form in $\nu$ and $\lambda$ which is is positive definite given Sylvester's criterions (\ref{sylvester:a}) and (\ref{sylvester:b}). This is a contradiction and (\ref{constraints}\,:c) follows. In addition to (\ref{sylvester:a})-(\ref{sylvester:c}) we are now equipped with positive $\varphi^2$ and $\gamma^2$. Next let us derive (\ref{CSconstraints}\,:b), that says $\vartheta^2 \gamma^2 - \nu^2>0$. By assuming $\nu = \pm \vartheta\gamma$ inequality (\ref{sylvester:c}) can be written $0 > (\gamma\mu \mp \vartheta\lambda)^2$ which is a contradiction and (\ref{CSconstraints}\,:b) follows. It only remains to derive (\ref{CSconstraints}\,:c), which can be proved by contradiction in a similar way by assuming $\lambda=\pm \varphi\gamma$.

\bibliographystyle{JHEP}

\bibliography{refs_ordered}

\end{document}